\documentclass[review]{elsarticle}

\usepackage{lineno,hyperref}
\modulolinenumbers[5]

\journal{Journal of \LaTeX\ Templates}
\usepackage{longtable}
\usepackage{float}
\floatstyle{plaintop}
\restylefloat{table}
\usepackage[tableposition=top]{caption}
\usepackage{graphicx}
\usepackage{amssymb}
\usepackage{multirow}
\usepackage{array}
\newcolumntype{P}[1]{>{\centering\arraybackslash}p{#1}}
\newcolumntype{M}[1]{>{\centering\arraybackslash}m{#1}}

\usepackage{amsmath}

\usepackage{caption} 
\usepackage{titlesec}
\titleformat{\paragraph}{\normalfont\normalsize\bfseries}{\theparagraph}{1em}{}
\titlespacing*{\paragraph}{0pt}{3.25ex plus 1ex minus .2ex}{1.5ex plus .2ex}

\usepackage{amsmath}

\begin{document}

\begin{frontmatter}

\title{DoubleU-Net++: Architecture with Exploit Multiscale Features for Vertebrae Segmentation}


\author[author1]{Simindokht Jahangard\corref{mycorrespondingauthor}}\cortext[mycorrespondingauthor]{Corresponding author}
\ead{s_jahangard@aut.ac.ir}
\author[author2]{Mahdi Bonyani}
\ead{m_bonyani96@ms.tabrizu.ac.ir}

\author[author3]{Abbas Khosravi}

\ead{abbas.khosravi@deakin.edu.au}


\address[author1]{Department of Robotics Engineering, Amirkabir University of Technology, Iran}
    
\address[author2]{Department of Computer Engineering\unskip, 
    University of Tabriz\unskip, Iran }

\address[author3]{Institute for Intelligent Systems Research and Innovation (IISRI), Deakin University}

\begin{abstract}
Accurate segmentation of the vertebra is an important prerequisite in various medical applications (E.g. tele surgery) to assist surgeons.  Following the successful development of deep neural networks, recent studies have focused on the essential rule of vertebral segmentation. Prior works contain a large number of parameters, and their segmentation is restricted to only one view. Inspired by DoubleU-Net, we propose a novel model named DoubleU-Net++ in which DensNet as feature extractor, special attention module from  Convolutional  Block  Attention  on  Module (CBAM) and,  Pyramid Squeeze Attention (PSA) module are employed to improve extracted features. We evaluate our proposed model on three different views (sagittal, coronal, and axial) of VerSe2020 and xVertSeg datasets. Compared with state-of-the-art studies, our architecture is trained faster and achieves higher precision, recall, and F1-score as evaluation (imporoved  by 4-6\%) and the result of above 94\%   for sagittal view and above 94\% for both coronal view and  above 93\% axial view were gained for VerSe2020 dataset, respectively. Also, for xVertSeg dataset, we achieved precision, recall ,and F1-score of above 97\% for sagittal view, above 93\% for coronal view ,and above 96\% for axial view.  
\end{abstract}


\end{frontmatter}


\section{Introduction}
The spine is the core of the skeleton and the skeleton is the core of the body. The spine has several functions such as support, stability, movement, shock absorption, and neurological integrity. The spine holds us up and transfers the weight of the torso to the pelvis, hips, and legs and all muscles attach to it. It is a highly  articulated structure made strong by immensely strong postural muscles which provide the body stability. locomotion would not be possible without good spine function and the natural curve of the spine, as well as the intervertebral discs, create a spring system allowing us to walk and run without immediately damaging ourselves. All the neurological signals traveling from the brain to the body are passing along this bony canal. Therefore, spinal problems, even subtle ones, such as the loss of the normal spinal curves, can affect any or all of these functions and the other parts of the body that depend upon them. The importance of this has led to extensive studies on understanding the biomechanics of spine structure \cite{loffler2020x, anitha2020effect, laouissat2018classification}. Despite the critical role of the spine, spinal pathology often reaches its critical final stages when treatment no longer affects the patient's recovery \cite{howlett2020radiology}. Therefore, identifying pathologies at an early stage presents an important opportunity to prevent and provide effective treatment. 

The computer-aided assistance facilitates the early diagnosis of spinal pathologies without human error. Fortunately, with the advent and rapid development of neural networks, especially deep networks, the analysis and understanding of medical images, especially the images of the spine to label and segment vertebrates has received special attention. Labeling and segmenting of vertebrate are two key components in analyzing and understanding spine image data because diagnosing spinal deformities including scoliosis, kyphosis, and lordosis, spinal curvature estimation, and vertebral fractures detection require vertebral labeling and vertebral segmentation.

Although vertebral labeling is a trivial task on a small dataset, it is a cumbersome and time-consuming task for large datasets. In addition, spine images usually have low resolution, and vertebra’s posterior elements have complex morphology that makes the delineation inaccurate and inconsistent. Traditionally, spine segmentation has been considered as a model-fitting problem using statistical shape models and its variants \cite{castro2015statistical,pereanez2015accurate,korez2015framework} or active contour \cite{athertya2016automatic,hammernik2015vertebrae}, while recent techniques focused on machine learning and deep learning (DL)-based methods\cite{lessmann2019iterative,kelm2013spine,sheng2021one,li2021whole}.

One of the limitations of the previous works is high computation time requiring multiple stages \cite{sekuboyina2021verse}. In contrast, the lightweight and minimum computational burden are the key advantages of the proposed solution. Using DoubleU-Net as our based network, we incorporated Convolutional Block Attention Module \cite{woo2018cbam} with Atrous Spatial Pyramid Pooling (ASPP) \cite{chen2017rethinking} and utilized Pyramid Squeeze Attention (PSA) module to improve extracted features. In addition, the unavailability of a dataset large enough is always a considerable problem; thus, we employed different augmentation to boost performance and avoid overfitting. We evaluated our proposed architecture on VerSe2020 and xVertSeg datasets which are the most significant challenges in the context of vertebrae segmentation and identification.

The rest of the paper is organized as follows. In Section 2, the existing literature and state-of-the-art in vertebral segmentation are concisely reviewed. In Section 3, the utilized dataset is presented, the proposed method is described from a technical perspective, including the details of the preprocessing and training procedures, along with the designed architecture. In Section 4, the results are presented and evaluated against the ones reported in the literature heretofore. Finally, in Section 5, the conclusions are listed, and suggestions on possible future research directions are put forward.

\section{Literature review}
Dong Zhang et al. \cite{zhang2021sequential} proposed a Sequential Conditional Reinforcement Learning network (SCRL) to tackle the problem of simultaneous detection and segmentation of the vertebral body. In their method, anatomical correlation of the spine was modeled by leveraging the decision-making process. Also, the new rewarding function is designed to improve accuracy by solving the Unbalanced-experience. 

U-Net-based models \cite{li2018h,chen2018s3d,moradi2019mfp,jahangard2020u,ma2021lf,du2021tsu} are extensively utilized in medical image segmentation. Dong Hyun Kim et al. \cite{kim2021automated} proposed multi dilated recurrent residual U-Net (MDR2-UNet) consisting of Multi Dilated Residual Block (MDRB) and Recurrent Residual Block (RRB). The model segmented the lateral spine X-ray images therefore Vertebral Compression Ratio (VCR) is measured with the segmented data. Faisal Rehman et al. \cite{rehman2020region}  presents a novel combination of region-based level set technique with the deep convolutional neural networks (CNNs). The region-based level set is based on the active contour model which is used to extract region-based information. Similarly, CNNs are also used in recent years for medical image segmentation. Sekuboyina et al.\cite{sekuboyina2017localisation} proposed a two-stage approach in which in the first stage a multi-layered perceptron was employed to localized the lumbar region and in the second stage a fully CNN was exploited to segment and label the lumbar vertebrae. They tested their method on a publicly available dataset of the xVertSeg segmentation challenge of MICCAI’16, achieving an average Dice coefficient of over 90\%. Janssens et al. \cite{janssens2018deep} utilized a model based on cascaded 3D Fully Convolutional Networks (FCNs), consisting of two parts: localization FCN and a segmentation FCN. In the first step the bounding box of the lumbar region was found, the network of this part named “LocalizationNet”. Then in the second part, a 3D U-net called “SegmentationNet” was developed to perform a pixel-wise multi-class segmentation, resulting in obtaining an average Dice coefficient of 95.77 ± 0.81\% on xVertSeg dataset. A single-stage network based on the MobileNetV3 and the DeepLab/ASPP was introduced by Hempe and Heinrich \cite{hempe2021towards}. Specifically, their model is comprised of a backbone model, the ASPP, and a segmentation head with a single skip connection to the backbone. They achieved Dice of 77.9\% on the VerSe20 Challenge dataset. Altini et al. \cite{altini2021segmentation} exploited deep learning and classical machine learning methodologies. In the first phase, using a 3D CNN, the whole spine was segmented in binary and traditional machine learning algorithms such as k-Means Clustering and k-NN were utilized for locating vertebrae centroids. Using the VerSe20 dataset, they achieved a binary Dice coefficient of 89.17\% and an average multi-class Dice coefficient of 90.09\%. Kim et al. \cite{kim2020web} utilized a method based on CNN. More specifically, a hierarchical data format file using U-Net architecture was trained and then the test data label was inserted to perform segmentation. They evaluated their own dataset, resulting in an average dice coefficient of 90.4\%, a precision of 96.81\%, and an F1-score of 91.64\%. Vania et al. \cite{vania2019automatic} combined CNN and fully convolutional networks (FCNs). Also, they used class redundancy as a soft constraint to greatly improve the segmentation results. They tested their result in terms of the Dice coefficient (94\%), Jaccard index (93\%), volumetric similarity (96\%), sensitivity (97\%), specificity (99\%), precision (over-segmentation 8.3 and under-segmentation 2.6), accuracy (99\%), Matthews correlation coefficient (0.93), mean surface distance (0.16 mm), Hausdorff distance (7.4 mm), and global consistency error (0.02). Qadri et al. \cite{furqan2019automatic} proposed a deep learning approach for automatic CT vertebra segmentation named patch-based deep belief networks (PaDBNs). In their method, the features were selected from image patches and then the differences between classes were measured. For weight initialization, an unsupervised feature reduction contrastive divergence algorithm is applied.

\section{Methods}
\subsection{Datasets}
In this study datasets: VerSe2020 dataset \cite{sekuboyina2021verse} and xVertSeg dataset \cite{sekuboyina2017localisation} were utilized.
\subsubsection{VerSe2020 dataset}
The VerSe2020 dataset consists of 319 CT scans and approximately 300 patients with a mean age of $\sim$59(±17) years participated to collect the dataset. The scans are split into 113, 103, and 103 for training, validation, and test set, respectively. The total number of vertebrae is 4141 in which cervical (CER) make up 581, thoracic (THO) consists of 2255, and lumbar (LUM) number is 1305. The dataset provides 3D data that we used 2D view sagittal and coronal planes for segmentation. Some samples of VerSe2020 dataset from sagittal, coronal and axial views are illustrated  in Figure 1.
\subsubsection{xVertSeg dataset}
The xVertSeg dataset was released in xVertSeg challenge of MICCAI 2016 \cite{sekuboyina2017localisation}. The dataset provides vertebral deformities, spine curvatures, and varying fields-of-views (FOVs). Fifteen train CT volumes with ground truth segmentation of the lumbar vertebrae (into five classes, L1-L5) and ten test CT volumes are included. The dataset consists of the ground truth for both localization and segmentations. In our study, our focus is on the segmentation part. Figure 2 depicts some samples of xVertSeg dataset from different views.

\begin{figure}[htp]
\centering
\includegraphics[width= 0.8\textwidth]{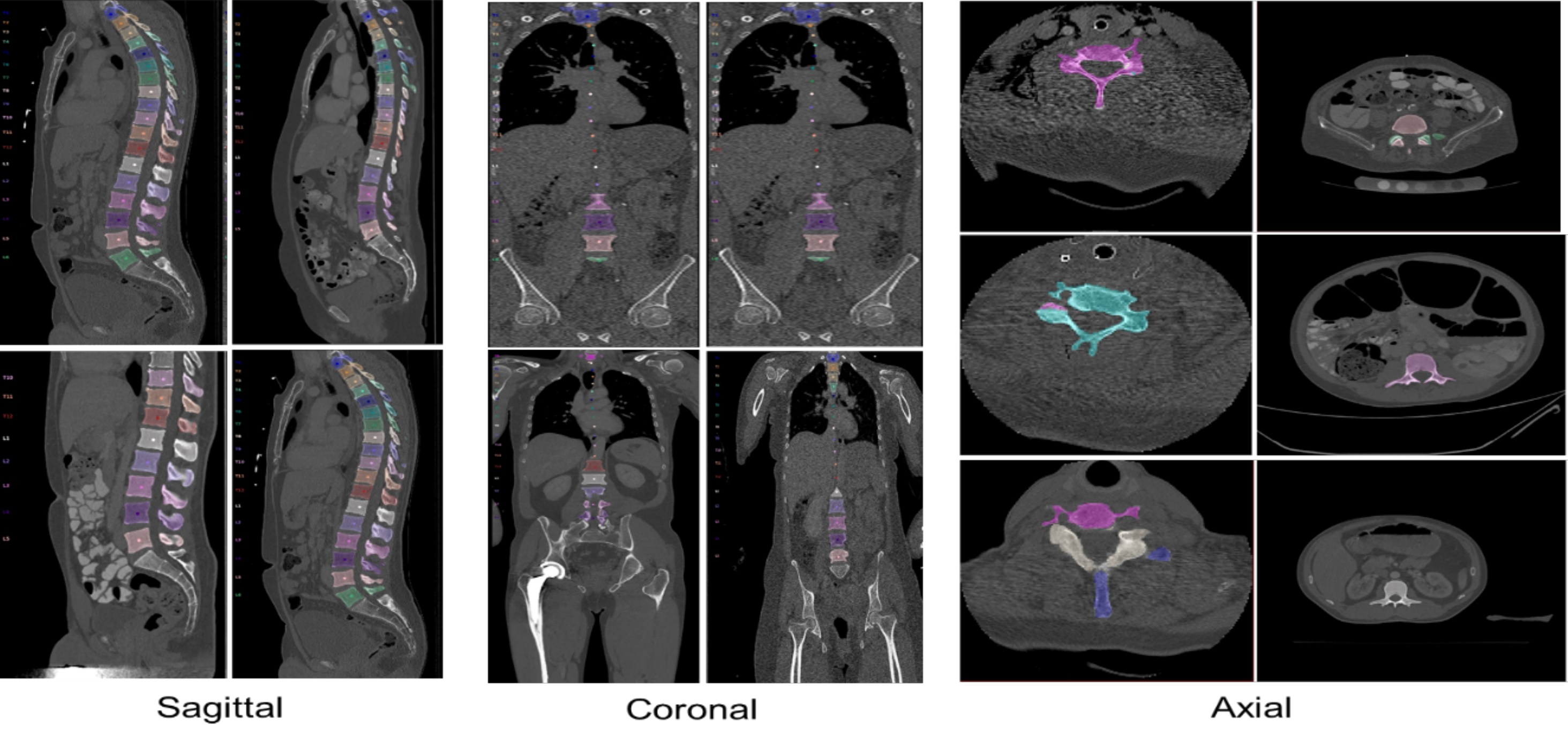}
\caption{\label{fig:fig1}Sample images from the VerSe2020  dataset from sagittal, coronal and, axial view.}
\end{figure}

\begin{figure}[htp]
\centering
\includegraphics[width= 0.8\textwidth]{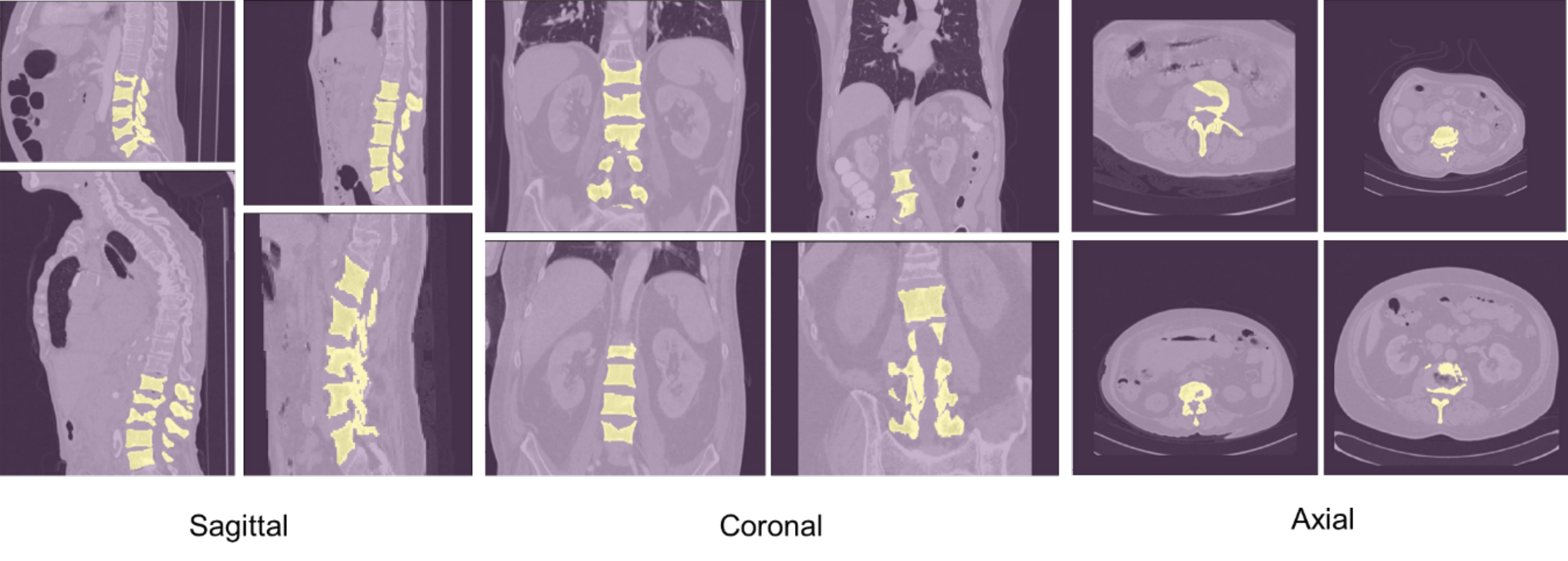}
\caption{\label{fig:fig1}Sample images from the xVertSeg dataset from sagittal, coronal and, axial view.}
\end{figure}

\subsection{Preprocessing}
Discriminating between the vertebrae and other tissues is the main goal in the preprocessing step. In this work, initially, we changed the voxel space value to one for both datasets. After changing the voxel space according to the domain of the intensity value in each of the CT images, we scale the intensity according to the following equation so that the intensity of pixels is normalized. The intensity of the pixels is divided by 2048, and then the value of each pixel is randomly shifted between -0.25 and 0.25, and in the next step, the value of each pixel is randomly scaled between -1.25 and 1.25,  and finally, the value of each pixel is clipped between -1 and 1. And after these steps, the input images to the network will be resized to 256 by 256 dimensions.

\subsubsection{Augmentation}
Data augmentation is also applied to alleviate the lack of training samples and avoid overfitting. We have used two sets of augmentation that are selected for each image with a probability of 0.6 either from the first set of augmentation or with a probability of 0.4 from the second set of augmentation. The first set of augmentation includes Flip\_left\_right, Flip\_up\_down, Central\_crop, Random\_crop, Random\_contrast, Random\_brightness, and Transpose. In the second set of augmentation, rotation, shear, zoom, and shift operations are included. The applied augmentation boosted the performance of both models significantly. The effect of augmentation on performance for both DoubleU-Net and the proposed model is provided in Section 4.

\subsection{Proposed Model}
\subsubsection{DoubleU-Net}
DoubleU-Net \cite{jha2020doubleu} is an encoder-decoder architecture for semantic image segmentation. It consists of two U-Net architectures stacked on top of each other. In the first U-Net, the VGG-19 \cite{simonyan2014very} as an encoder has been adopted which is trained on ImageNet \cite{deng2009imagenet} that learned features already to be usable and ready for other tasks. DoubleU-Net utilized VGG-19 network because compared with other pretrained networks, it is very lightweight and its architecture is similar to U-Net, resulting in easier concatenation. Moreover, it provides a deeper network, enabling a more accurate segmentation mask. The second U-Net is added at the bottom of the first U-net to leverage more semantic information efficiently. DoubleU-Net is able to capture contextual information within the network serving Atrous Spatial Pyramid Pooling (ASPP) \cite{chen2017rethinking}. DoubleU-Net outperformed other segmentation algorithms on four datasets: 2015 MICCAI sub-challenge on automatic polyp detection dataset, 2018 Data Science Bowl challenge dataset, Lesion Boundary Segmentation challenge from ISIC-2018, and CVC-ClinicDB dataset. Therefore, we selected DoubleU-Net as our basic architecture. 
\subsubsection{DoubleU-Net++}
In our model, the DenseNet is employed instead of VGG-19 in architecture. As their main advantage in comparison to VGG-19, while still being deep enough to capture the distinctive nuances based on an efficient selection of features. Furthermore, it has significantly fewer parameters, which makes them more suitable, in terms of avoiding overfitting, on the basis of the existing relatively small datasets of vertebral datasets. DenseNet usually has 7,031,232 parameters, being significantly smaller than the number of parameters of VGG-19 architects, i.e. about 20,023,232 parameters. Similarly, our proposed model, DoubleU-Net++, has 15,924,886 parameters, while DoubleU-Net has 29,297,570 parameters.

The reason behind utilizing ASPP in the output of the encoder, DensNet, is to enhance the quality of output features. ASPP has five convolution layers in which the outputs of these convolutions are concatenated. However, in our model, we exploit Convolutional Block Attention Module (CBAM) \cite{woo2018cbam} after ASPP to boost the quality of output features. Although CBAM includes two distinctive modules, channel and spatial, we employed only spatial attention. For spatial attention average-pooling and max-pooling operations were applied and then concatenated to generate an efficient feature descriptor to boost the performance. Using spatial attention in CBAM not only refines the features, resulting in higher performance but also the overall overhead of CBAM is quite small in terms of both parameters and computation. Furthermore, the attention of the model instead of focusing on only a single vertebrate will focus on the whole spine which results in finer segmentation in the next modules. The spatial attention equation is shown in (1).
\begin{equation}
  F'={M{_s}(F)}\otimes  F
\end{equation}
$ F'\in R^{C \times H \times W} $ is input feature map and $M{_s}\in R^{1 \times H \times W} $ is a 2D spatial attention map and  \( \otimes \) denotes element-wise multiplication.

We use Pyramid Squeeze Attention (PSA) module \cite{zhang2021epsanet} which is a low-cost and high-performance novel module. PSA modules are able to integrate the information of the input feature map using the multi-scale pyramid convolution structure. The spatial information with different scales from each channel-wise feature map is effectively extracted by squeezing the channel dimension of the input tensor. This enables neighbor scales of context features to be more precisely incorporated. Then, a cross-dimension interaction is built by extracting the channel-wise attention weight of the multi-scale feature maps.

In network 2, in the decoder block, the two layers of convolution have been utilized followed by batch normalization and Relu as activation function. Finally, in the last layer, the squeeze block is employed acting like attention. In the squeeze block used in DoubleU-Net, initially, the input is flattening and a layer of Dense is used to determine which feature is important. It is then multiplied by the input to extract the important features. In our model, we added Random Feature \cite{rahimi2007random} in squeeze block. Specifically, the random feature alters the space of features which improves the low level features . The random feature imposes many advantages. First, as it is random, it plays a role like augmentation on low-level features in the last layers of the model. Therefore, the model is able to have a better understanding of features.

\begin{figure}[htp]
\centering
\includegraphics[width= 0.8\textwidth]{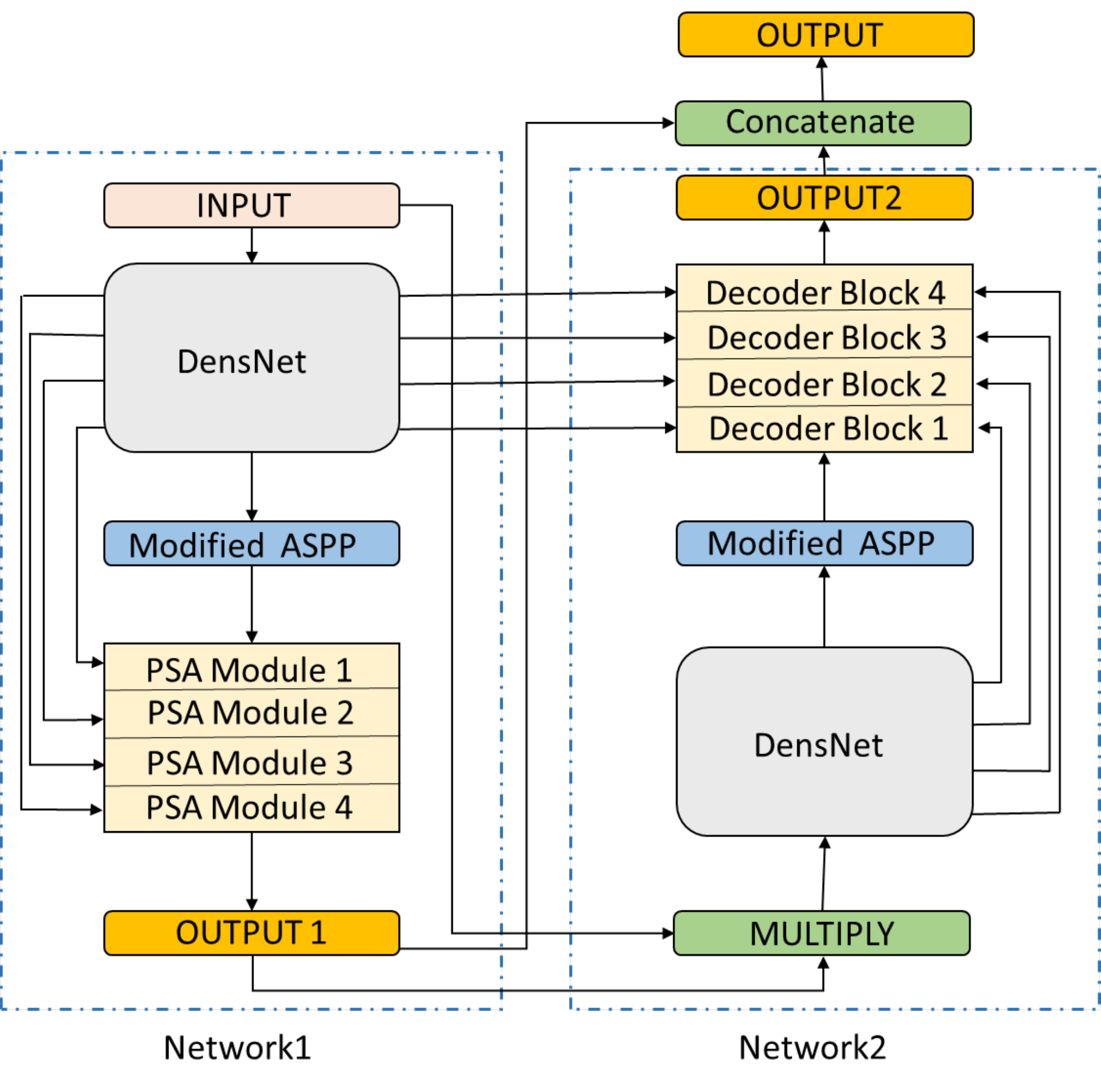}
\caption{\label{fig:fig3}Block diagram of the proposed DoubleU-Net++ architecture}
\end{figure}

\subsubsection{Training}
To train our network, the number of epochs is 160 with a batch size of 8, a custom learning rate (1e-05 to 0.00048 to 0.000152), and an Adam \cite{kingma2014adam} optimizer. Figure 4 shows the trend of learning rate during 160 epochs. The network was trained using the binary cross-entropy along with the Dice loss function and the training process was conducted in Google Colab graphics processing unit. The learning curve of the DoubleU-Net and DoubleU-Net++ for F1-score for both datasets have been depicted in Figure 5 and Figure 6, respectively.

\begin{figure}[htp]
\centering
\includegraphics[width= 0.8\textwidth]{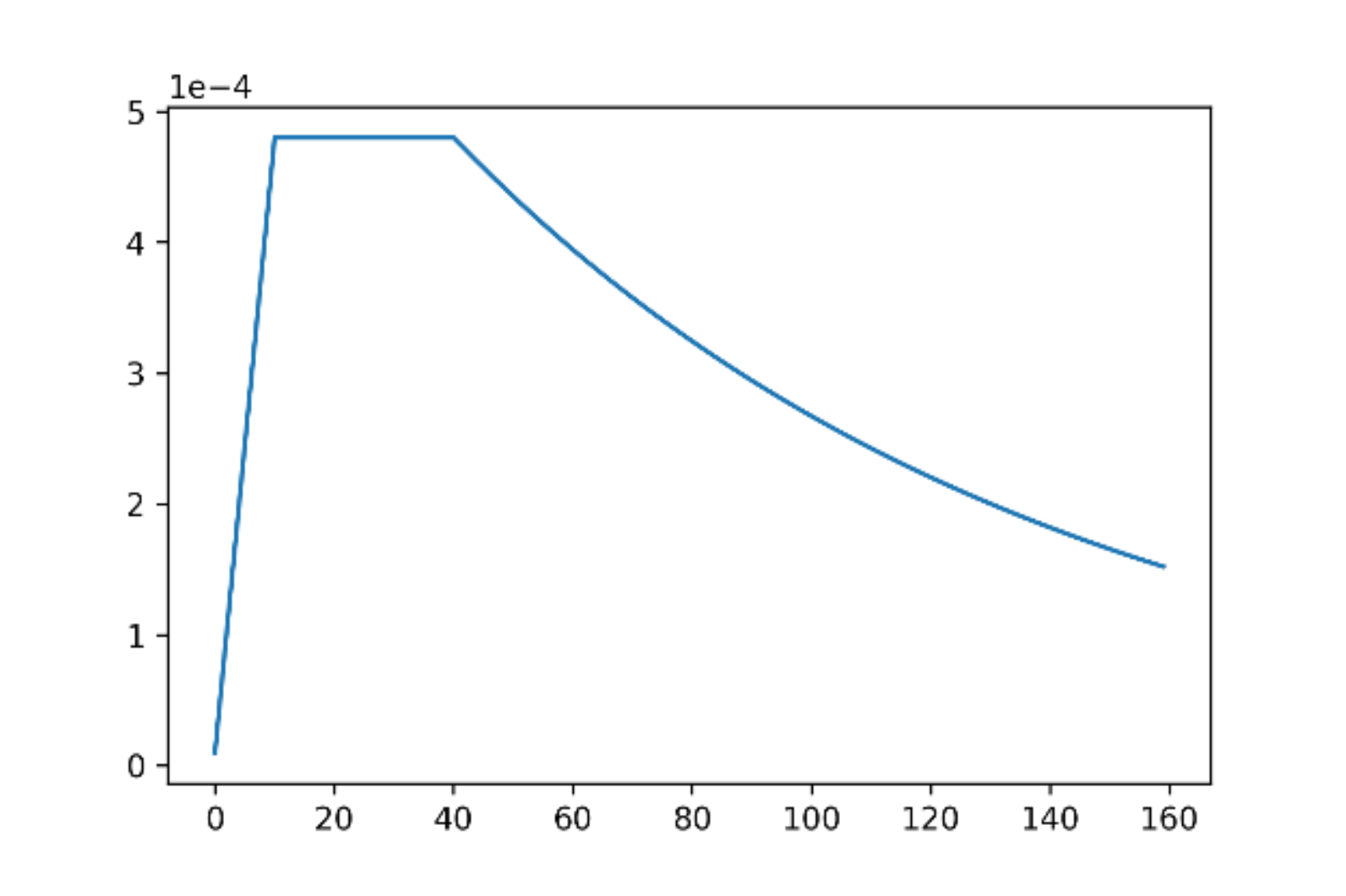}
\caption{\label{fig:fig4}Custom learning rate during the training process.}
\end{figure}

\begin{figure}[htp]
\centering
\includegraphics[width= 0.8\textwidth]{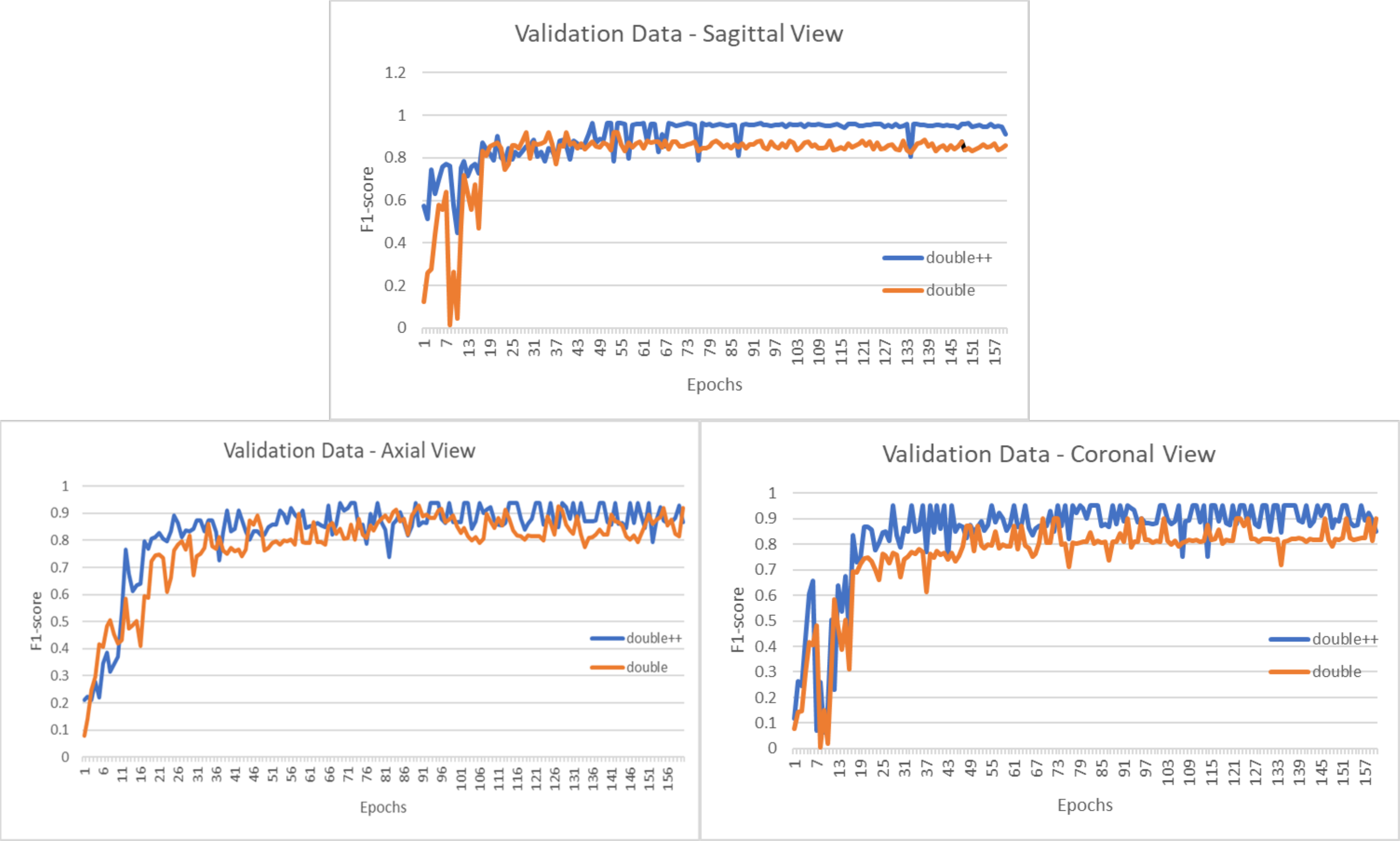}
\caption{\label{fig:fig5}The process of training for DoubleU-Net and DoubleU-Net++ on VerSe2020 dataset for F1-score.}
\end{figure}

\begin{figure}[htp]
\centering
\includegraphics[width= 0.8\textwidth]{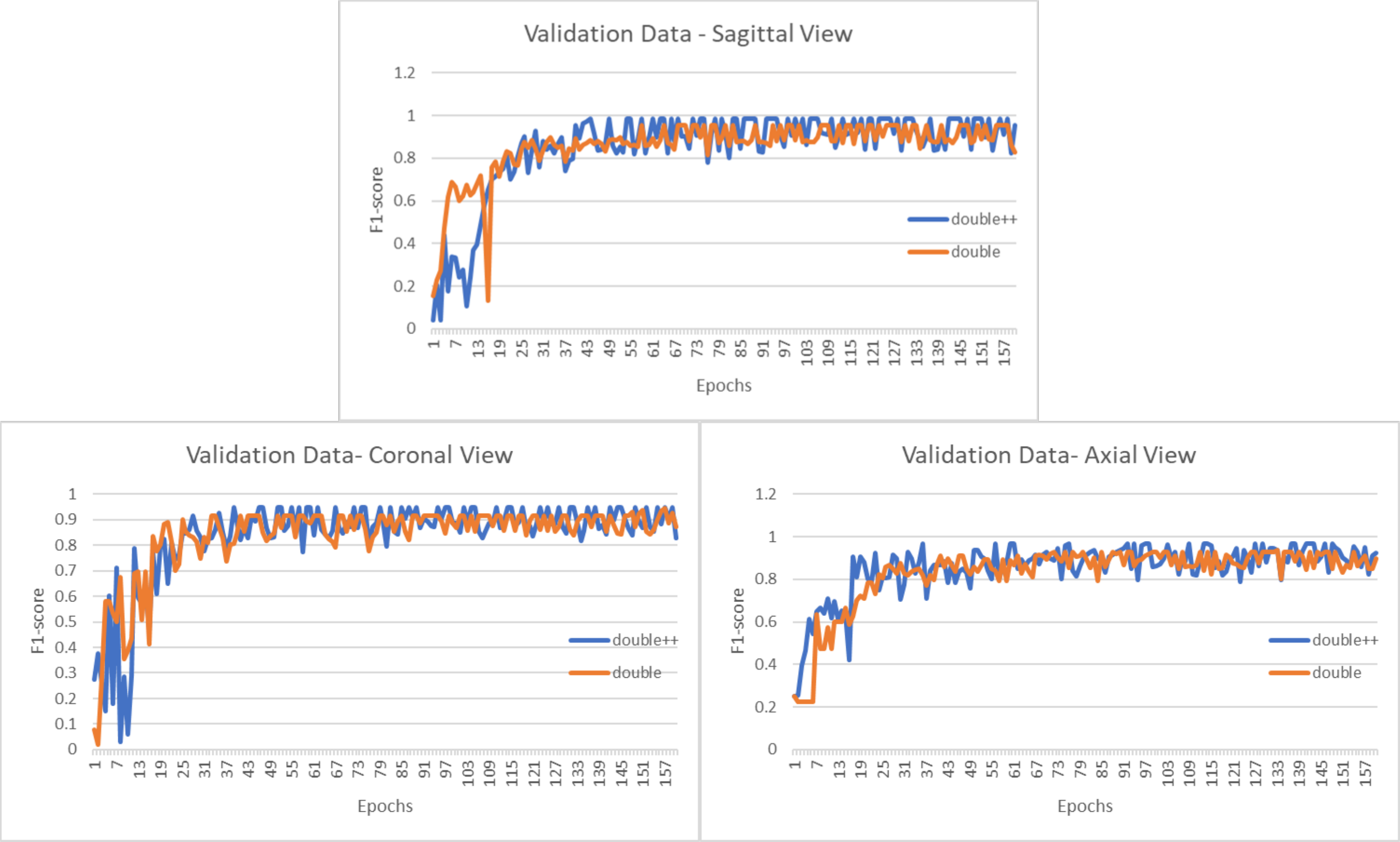}
\caption{\label{fig:fig6}The process of training for DoubleU-Net and DoubleU-Net++ on xVertSeg dataset for F1-score.}
\end{figure}
\subsubsection {Network evaluation}
To evaluate the performance of the proposed model, we use three standard performance metrics, described in Eq (1) to (3): precision, recall, F1-score. There is a point that The F1-score is also known as the Dice coefficient or Dice similarity coefficient (DSC). The elements to calculate the mentioned metrics are True Positive (TP), True Negative (TN), False Positive (FP), and False Negative (FN), that are defined as: 

•	True Positive (TP) = pixels correctly segmented as foreground (vertebrae)

•	False Positive (FP) = pixels falsely segmented as foreground (vertebrae)

•	True Negative (TN) = pixels correctly detected as background

•	False Negative (FN) = pixels falsely detected as background

\section{Results and Discussion}
In this section, the performance of DoubleU-Net and DoubleU-Net++ for VerSe2020 and xVertSeg datasets are listed in Table1 and Table2, respectively. The comparison is made from three views: sagittal, coronal, and axial in terms of precision, recall, and F1-score metrics for both valid and test data. In Table1, for sagittal view, our model, which resulted in a precision of 96.34\%, recall of 95.94\%, and F1-score of 96.13\% for valid data and precision of 95.18\%, recall of 94.08\%, and F1-score of 94.62\% for test data, outperforms DoubleU-Net. Similarly, for the coronal view, we achieved a precision of 95.12\%, recall of 95.43\%, and F1-score of 95.27\% for valid data and precision of 94.02\%, recall of 94.24\%, and F1-score of 94.12\% for test data, while these evaluation metrics for DoubleU-Net are approximately by 4\% lower. Regarding axial view, the precision, recall, and F1-score of 94.69\%, 94.93\%, and 94.80\% for valid data and 94.01\%, 93.86\%, and 93.93\% for test data are obtained, which is significantly improved in comparison with DoubleU-Net. 
\begin{center}

\begin{table}[t!]
\caption{\centering The comparison of precision, recall and F1-score for our model and Double-net in percentage from three different views for VerSe 2020 dataset.}
    \label{table:1}
\resizebox{1.\textwidth}{!}{%
\begin{tabular}{ccccccc}
\hline

\textbf{Plane} & \textbf{Model}& \textbf{Phase }& \textbf{Precision(\%)} &\textbf{Recall(\%)}&\textbf{F1-score(\%)} \\\hline

\multirow{4}{*}{Sagittal } & \multirow{2}{*}{DoubleU-Net } &Valid& 92.19   & 91.49 & 91.83  \\  & & Test  & 90.23  & 90.14 & 90.18 \\\cline{2-6}
& \multirow{2}{*}{\textbf{DoubleU-Net ++} } &Valid& \textbf{96.34}    & \textbf{95.94}  & \textbf{96.13}   \\     & & Test  & \textbf{95.18}   & \textbf{94.08}  & \textbf{94.62}  \\  
\hline

\multirow{4}{*}{Coronal } & \multirow{2}{*}{DoubleU-Net } &Valid& 90.09   & 90.13 & 90.10  \\  & & Test  & 89.51  & 89.04 & 89.27  \\\cline{2-6}
& \multirow{2}{*}{\textbf{DoubleU-Net ++ }} &Valid&\textbf{95.12}   &\textbf{95.43} & \textbf{95.27}  \\  & & Test  & \textbf{94.02}  & \textbf{94.24} &\textbf{94.12 } \\  \hline
\multirow{4}{*}{Axial } & \multirow{2}{*}{DoubleU-Net } &Valid& 92.37   & 92.51 & 92.43  \\  & & Test  & 91.45  & 91.73 & 91.58  \\\cline{2-6}
& \multirow{2}{*}{\textbf{DoubleU-Net ++}} &Valid& \textbf{94.69}  & \textbf{94.93} & \textbf{94.80}  \\  & & Test  & \textbf{94.01} & \textbf{93.86}& \textbf{93.93}  \\                       
\hline

\end{tabular}
}
\end{table}
\end{center}

Moreover, our proposed method is tested on xVertSeg dataset, which resulted in desirable precision of 98.49\% and 97.95\%, recall of 98.23\% and 97.54\%, and F1-score of 98.35\% and 97.74\% for valid and test data for the sagittal view, as illustrated in Table2. Furthermore, as it can be illustrated, for coronal view, DoubleU-Net++ could surpass DoubleU-Net, in which for precision, recall, and F1-score of 94.48\%, 94.83\% and, 94.65\% was achieved for valid data and 93.79\%, 93.64\%, and 93.71\% were gained for test data. When it comes to axial view, again, it can be seen that the precision of 96.49\% and 96.02\%, recall of 96.83\% and 96.17\%, and F1-score of 96.65\% and 96.09\% for valid and test data achieved in this paper surpass the ones obtained in DoubleU-Net.

Moreover, the segmentation results generated by DoubleU-Net and DoubleU-Net++ are provided in Figure 7 and Figure 8 for VerSe2020 and xVertSeg datasets, respectively. Each row depicts an image on sagittal view. The first and second columns show the segmented image, described in Section 3.3.2 (Figure 3), and the third and fourth columns illustrate the final output of DoubleU-Net and DoubleU-Net++, respectively. Compared with DoubleU-Net the final outputs of DoubleU-Net++are more close to the ground truth which is shown in the last column.

\begin{center}

\begin{table}[t!]
\caption{\centering The comparison of precision, recall and F1-score for our model and Double-net in percentage from three different views for xVertSeg dataset.}
    \label{table:2}
\resizebox{1.\textwidth}{!}{%
\begin{tabular}{ccccccc}
\hline
\textbf{Plane} & \textbf{Model}& \textbf{Phase }& \textbf{Precision(\%)} &\textbf{Recall(\%)}&\textbf{F1-score(\%)} \\\hline

\multirow{4}{*}{Sagittal} & \multirow{2}{*}{DoubleU-Net } &Valid& 93.14  & 93.49 & 93.31 \\  & & Test  & 92.23  & 92.94 & 92.58 \\\cline{2-6}
& \multirow{2}{*}{\textbf{DoubleU-Net ++} } &Valid& \textbf{98.49}    & \textbf{98.23}  & \textbf{98.35}   \\     & & Test  & \textbf{97.95}   & \textbf{97.54}  & \textbf{97.74}  \\  
\hline

\multirow{4}{*}{Coronal } & \multirow{2}{*}{DoubleU-Net } &Valid& 91.48   & 91.34 & 91.40 \\  & & Test  & 90.74  & 90.01 & 90.37  \\\cline{2-6}
& \multirow{2}{*}{\textbf{DoubleU-Net ++ }} &Valid&\textbf{94.48}   &\textbf{94.83} & \textbf{94.65}  \\  & & Test  & \textbf{93.79}  & \textbf{93.64} &\textbf{93.71} \\  \hline
\multirow{4}{*}{Axial } & \multirow{2}{*}{DoubleU-Net } &Valid& 92.81  & 92.94 & 92.87 \\  & & Test  & 92.16  & 92.11 & 92.13  \\\cline{2-6}
& \multirow{2}{*}{\textbf{DoubleU-Net ++}} &Valid& \textbf{96.49}  & \textbf{96.83} & \textbf{96.65}  \\  & & Test  & \textbf{96.02 } & \textbf{96.17}& \textbf{96.09}  \\                       
\hline

\end{tabular}
}
\end{table}
\end{center}

\begin{figure}[htp]
\centering
\includegraphics[width= 0.8\textwidth]{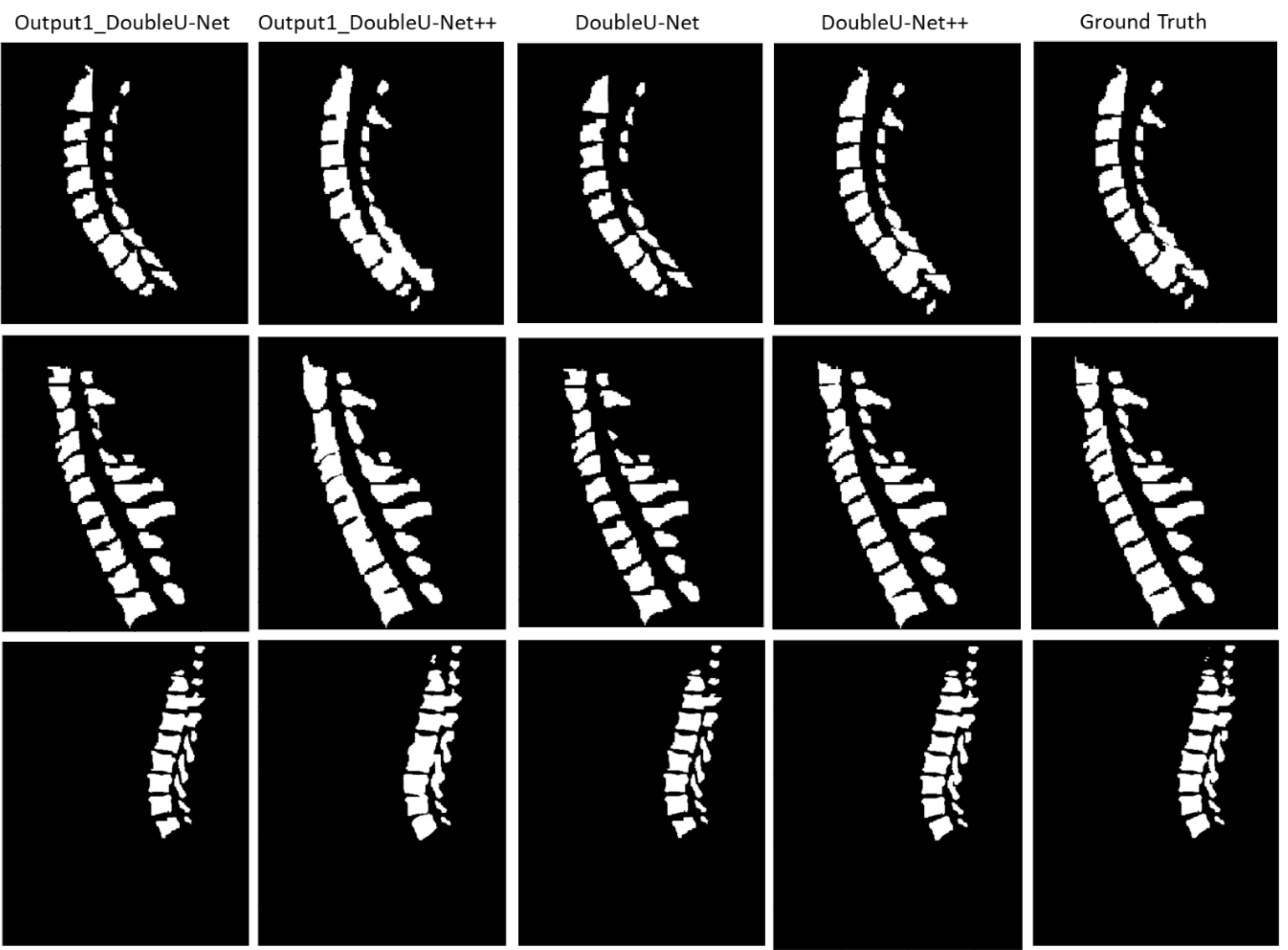}
\caption{\label{fig:fig7}Qualitative segmentation results for the VerSe2020 dataset on three sample images. Each row shows the results for one image. For each image, the ground truth segmentations are indicated in the last column and the output of Network1 and Network2 of DoubleU-Net and DoubleU-Net++ segmentations are exposed in the first and second, and in the third and fourth columns, respectively.}
\vspace*{\floatsep}%
\includegraphics[width= 0.8\textwidth]{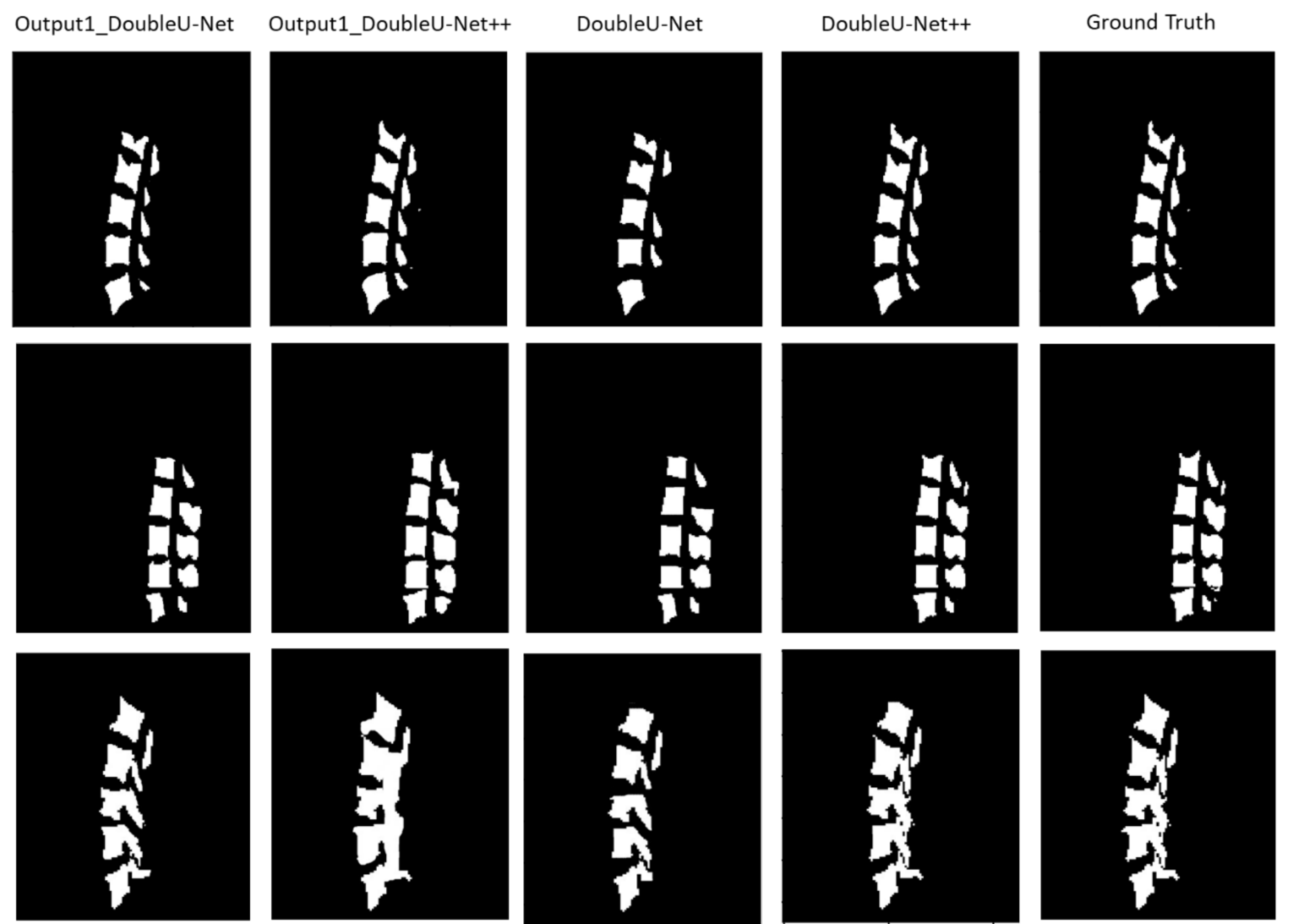}
\caption{\label{fig:fig1}Qualitative segmentation results for the xVertSeg dataset on three sample images. Each row shows the results for one image. For each image, the ground truth segmentations are indicated in the last column and the output of Network1 and Network2 of DoubleU-Net and DoubleU-Net++ segmentations are exposed in the first and second, and in the third and fourth columns, respectively.}
\end{figure}

In Table 3, we compare obtained results in terms of F1-score (dice), precision, and recall with state-of-the-art works. As it can be seen, for VerSe2020 dataset, our method with F1-score of 94.62\%, precision of 95.18\%, and recall of 94.08\% outperformed other studies worked on this dataset. Our results are from sagittal, coronal, and axial views provided in Table1 and Table 2. However, to compared with other works, the suitable plane is presented in Table 3. For xVertSeg dataset also, our model with dice, precision, and recall of 97.74\%, 97.95\% and 97.54\% surpassed alternative studies. Some other papers evaluating their methods on private datasets are also provided in Table 3.

\begin{center}

\begin{table}[t!]
\caption{\centering Comparison of our method with the state-of-the-art alternatives in terms of precision, recall and F1-score (dice), based on for different datasets.}
    \label{table:3}
\resizebox{1.\textwidth}{!}{%
\begin{tabular}{lcccc}
\hline
\textbf{References} & \textbf{Dataset}& \textbf{(F1-score)Dice(\%) }& \textbf{Precision/Recall(\%)} &\textbf{Plane} \\\hline

Hempe and Heinrich \cite{hempe2021towards} & VerSe2020 &77.9& N/A  & Sagittal \\
Altini et al. \cite{altini2021segmentation} & VerSe2020 &89.17& 85.43 $\pm$ 2.75/ 94.51$\pm$3.31  & Sagittal \\
Payer et al. \cite{payer2020coarse} & VerSe2020 &94 + 0.11& N/A  & Sagittal \\
\textbf{Our Model} & \textbf{VerSe2020} &\textbf{94.62}& \textbf{95.18/94.08}  & \textbf{Sagittal} \\
Sekuboyina et al. \cite{sekuboyina2017localisation} & xVertSeg &94.3$\pm$2.8& N/A  & Sagittal \\
Janssens et al. \cite{janssens2018deep}  & xVertSeg &95.77$\pm$0.81& N/A  & Sagittal \\
\textbf{Our Model} & \textbf{xVertSeg} &\textbf{97.74}& \textbf{97.95/97.54} & \textbf{Sagittal} \\
Qadri et al. \cite{furqan2019automatic} & 2nd MICCAI workshop data &86.1& N/A  & Axial \\
Kim et al. \cite{kim2020web}  & Own dataset &90.4& 96.81/-  & Axial \\
Vania et al. \cite{vania2019automatic}  & Own dataset &94& N/A  & Axial \\
Zareie et al. \cite{zareie2018automatic}  & Own dataset & 95.0$\pm$2.3 & N/A  & Axial \\
\textbf{Our Model} & \textbf{VerSe2020} &\textbf{93.93}& \textbf{94.01/93.86} & \textbf{Axial} \\
\textbf{Our Model} & \textbf{xVertSeg} &\textbf{96.09}& \textbf{96.02/96.17} & \textbf{Axial} \\
\hline

\end{tabular}
}
\end{table}
\end{center}

\subsection{Augmentation}
As described in Section 3.2.1, the data augmentation is applied to the training data to alleviate the lack of training samples and improve the performance. The comparison of models, DoubleU-Net and DoubleU-Net++, with and without augmentation is provided in Figure 9, Figure 10, Figure 11, and Figure 12 for all three views, sagittal, coronal, and axial. As it can be seen from Figure 9 and Figure 10, the performance increases at least 2\% in all evaluation metrics for both validation and test samples of both datasets. Similarly, the effect of augmentation shows significant improvement as shown in Figure 11 and Figure 13. 
\begin{figure}[htp]
\centering
\includegraphics[width= 0.8\textwidth]{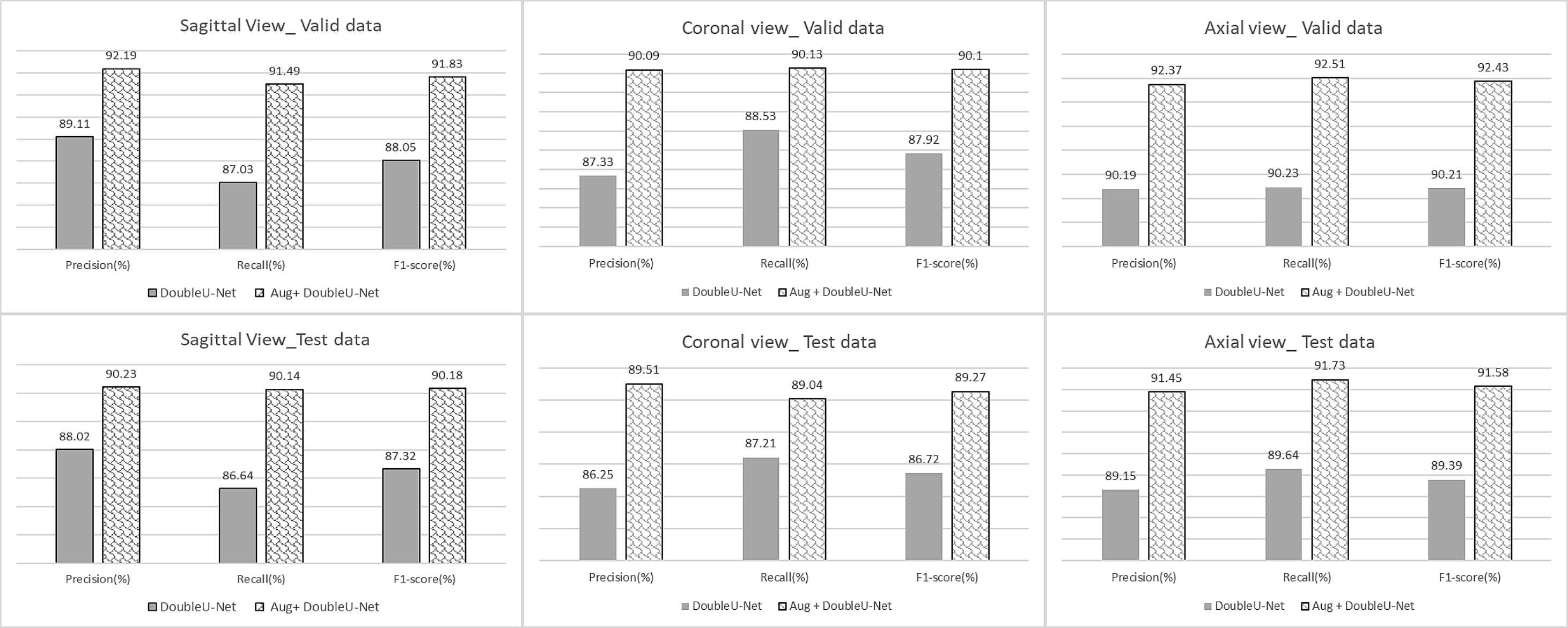}
\caption{\label{fig:fig9}The effect of augmentation on DoubleU-Net on performance for VerSe2020 dataset for both valid and test data}
\end{figure}
\begin{figure}[htp]
\centering
\includegraphics[width= 0.8\textwidth]{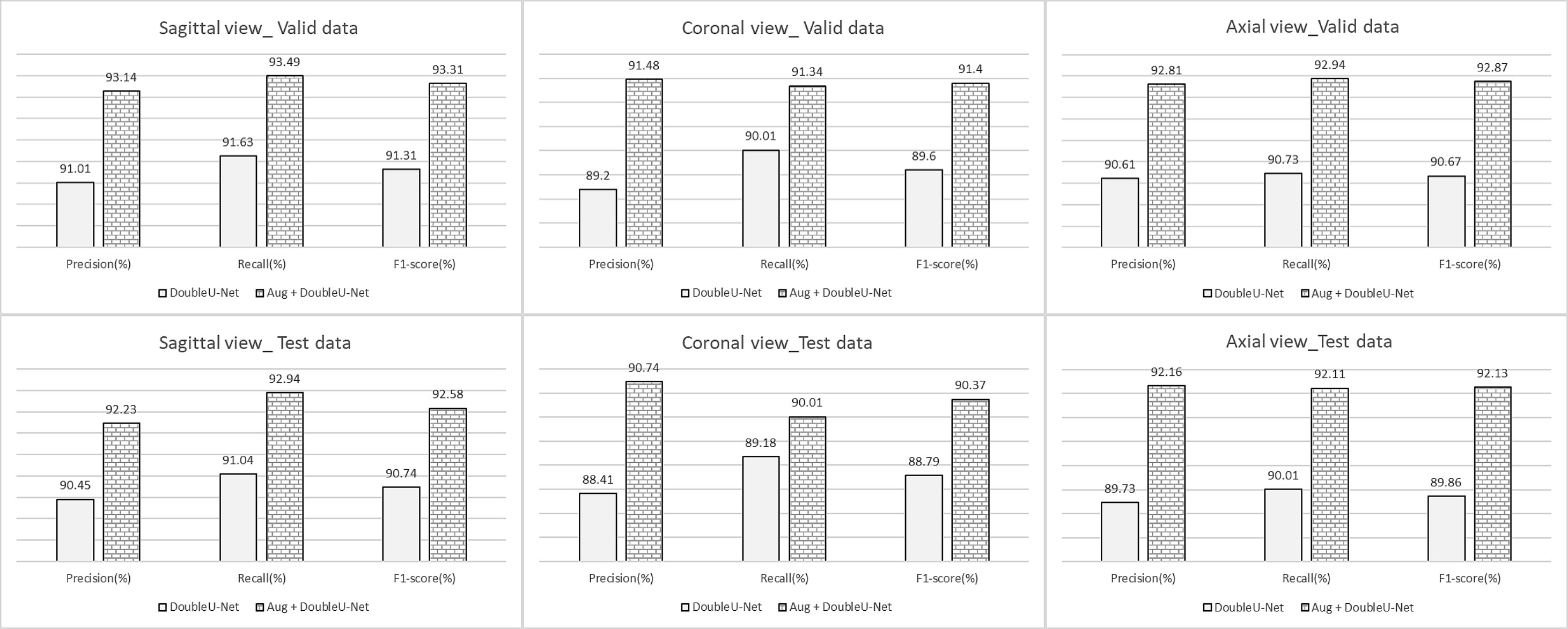}
\caption{\label{fig:fig10}The effect of augmentation on DoubleU-Net on performance for xVertSeg dataset for both valid and test data}
\end{figure}
\begin{figure}[htp]
\centering
\includegraphics[width= 0.8\textwidth]{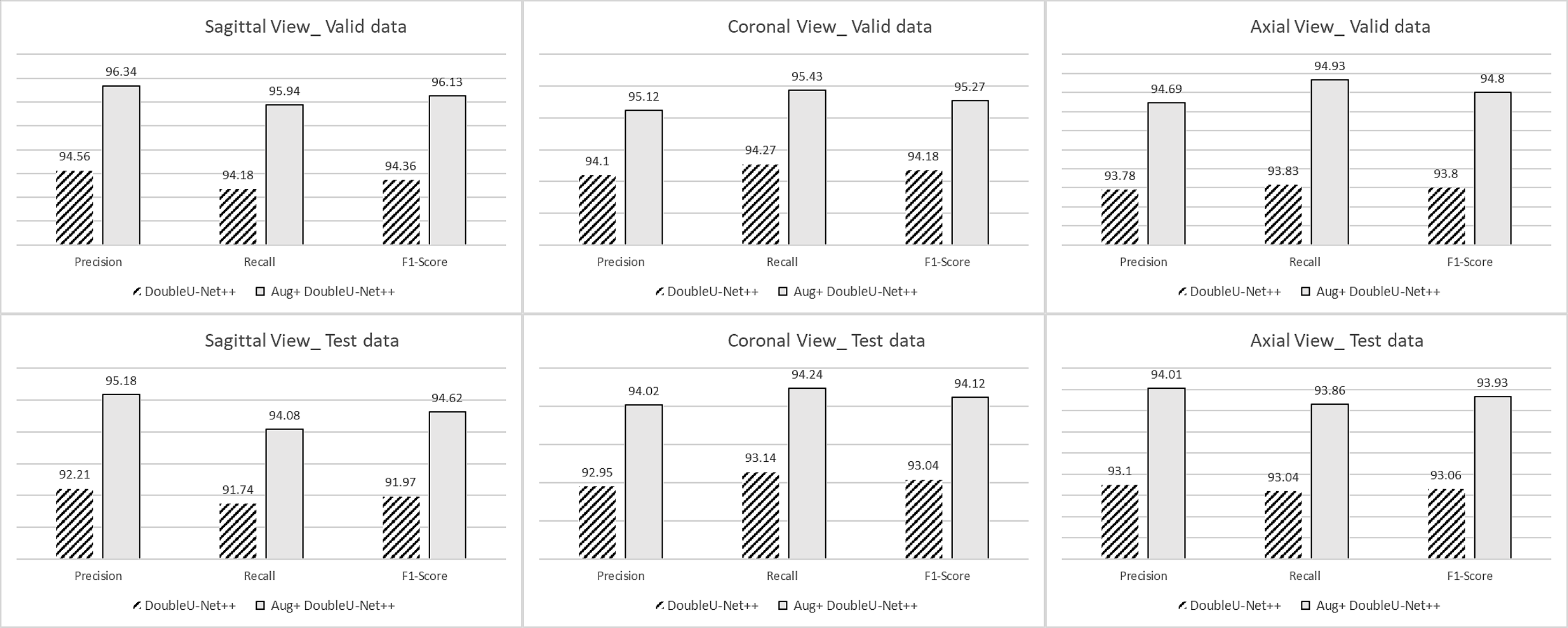}
\caption{\label{fig:fig11}The effect of augmentation on DoubleU-Net++ on performance for VerSe2020 dataset for both valid and test data}
\end{figure}
\begin{figure}[ht!]
\centering
\includegraphics[width= 0.8\textwidth]{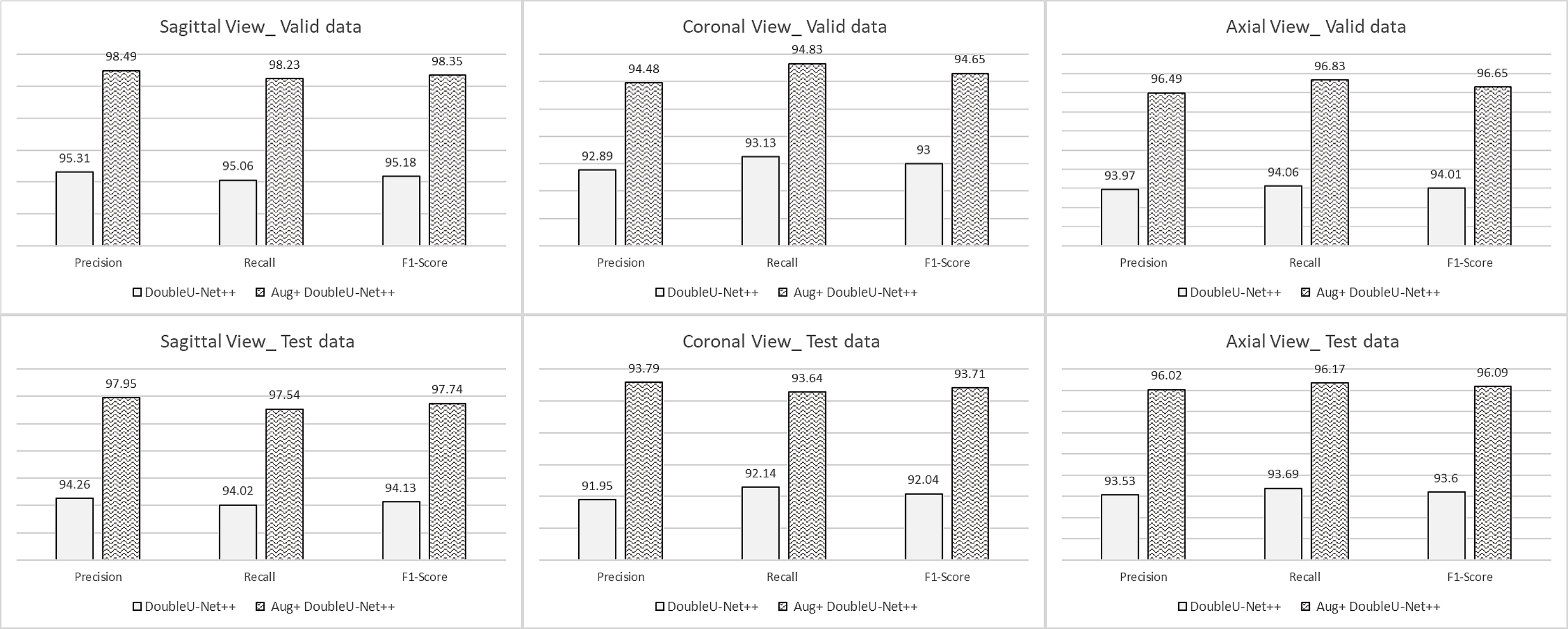}
\caption{\label{fig:fig12}The effect of augmentation on DoubleU-Net++ on performance for xVertSeg dataset for both valid and test data}
\end{figure}
\section{Conclusion}
The foundations for spine analysis such as fractures, malfunctions are the accurate segmentation and identification of vertebrae. The two large-scale vertebrae segmentation datasets (VerSe2020 and xVertSeg) are aimed at vertebrae segmentation and labeling. In this study, a novel deep neural network is proposed to segment vertebral medical images. The proposed solution is inspired by DoubleU-Net architecture in which the modified ASPP in which CBAM is used, PSA, and random feature contribute to more robust labeling results. We exploited robust augmentation, which had an efficient contribution to the final result and examined the model in sagittal, coronal, and axial views. Unlike other approaches, the proposed method comes with the added advantage of a lightweight network and less computational time. The obtained results illustrate that our DoubleU-Net++ outperforms the DoubleU-Net and other networks while having a lower number of parameters and training time. The obtained results for both VerSe2020 and xVertSeg datasets including precision, recall, F1-score (Dice) surpassed alternative methods. 


\bibliography{driverbibfile}

\end{document}